\documentclass{aastex}

\input psfig 
\def\puncspace{\ifmmode\,\else{\ifcat.\C{\if.\C\else\if,\C\else\if?\C\else%
\if:\C\else\if;\C\else\if-\C\else\if)\C\else\if/\C\else\if]\C\else\if'\C%
\else\space\fi\fi\fi\fi\fi\fi\fi\fi\fi\fi}%
\else\if\empty\C\else\if\space\C\else\space\fi\fi\fi}\fi}
\def\SP{\let\\=\empty\futurelet\C\puncspace }
\def\etal{et\SP al.\SP }

\def\kms{kms$^{-1}$}
\def\h-1{$h^{-1}$}

\def\void#1{{}}

\def\h1{$h^{-1}$}

\def\kms{kms$^{-1}$ }
\def\etal{et al.\,}

\def\eg{e.g., \,}

\def\lsim{~\rlap{$<$}{\lower 1.0ex\hbox{$\sim$}}}
\def\gsim{~\rlap{$>$}{\lower 1.0ex\hbox{$\sim$}}}

\def\mg{Mg$_2$\SP}

\begin{document}

\title{Redshift-Distance Survey of Early-Type Galaxies: Spectroscopic Data}


\author{
G. Wegner\altaffilmark{\ref{Hanover}},
M. Bernardi\altaffilmark{\ref{CMU}},
C. N. A. Willmer\altaffilmark{\ref{UCSC},\ref{Rio},\ref{onleave}},
L. N. da Costa\altaffilmark{\ref{ESO},\ref{Rio},\ref{onleave}},
M. V. Alonso\altaffilmark{\ref{AR},\ref{FR}},
P. S. Pellegrini\altaffilmark{\ref{Rio},\ref{onleave}},
M. A. G. Maia\altaffilmark{\ref{Rio},\ref{onleave}},
O. L. Chaves\altaffilmark{\ref{onleave}},
C. Rit\'e\altaffilmark{\ref{ESO},\ref{Rio},\ref{onleave}}
}

\newcounter{address}
\setcounter{address}{1}
\altaffiltext{\theaddress}{\stepcounter{address}
Department of Physics \& Astronomy, Dartmouth College, 6127 Wilder
Laboratory, Hanover, NH 03755, U.S.A.\label{Hanover}}

\altaffiltext{\theaddress}{\stepcounter{address}
Department of Physics, Carnegie Mellon University, Pittsburgh, PA 15213
\label{CMU}}

\altaffiltext{\theaddress}{\stepcounter{address}
UCO/Lick Observatory, University of California,
1156 High Street, Santa Cruz,  CA 95064, USA\label{UCSC}}

\altaffiltext{\theaddress}{\stepcounter{address}
European Southern Observatory,
Karl-Schwarzschild Strasse 2, D-85748 Garching, Germany\label{ESO}}

\altaffiltext{\theaddress}{\stepcounter{address}
Observatorio Astr\'onomico de
C\'ordoba,  Laprida  854, C\'ordoba, 5000, Argentina and CONICET, Argentina
\label{AR}}

\altaffiltext{\theaddress}{\stepcounter{address}
CNRS UMR 5572, Observatoire Midi-Pyr\'en\'ees, 14 Avenue E. Belin, 31400 
Toulouse, France \label{FR}}

\altaffiltext{\theaddress}{\stepcounter{address}
Observat\'orio do Valongo, Ladeira do Pedro Antonio 43, Rio de Janeiro, R. J.,20080-090, Brazil\label{Rio}}

\altaffiltext{\theaddress}{\stepcounter{address}
Observat\'orio Nacional, Rua General Jos\'e Cristino, 77,
Rio de Janeiro, RJ 20921-400, Brazil
\label{onleave}}

\begin{abstract}
 
We present central velocity dispersions and Mg$_2$ line indices
for an all-sky sample of $\sim$ 1178  elliptical and S0 galaxies, of which 
984 had no previous measures.
This sample contains the largest set of homogeneous spectroscopic data
for a uniform sample of elliptical galaxies in the nearby universe.
These galaxies were observed as part of the ENEAR project, designed to
study the peculiar motions and internal properties of the local
early-type galaxies. Using 523 repeated observations of 317 
galaxies obtained during different runs, the data are brought 
to a common zero-point.  These multiple observations, taken during the
many runs and different instrumental setups employed for this project,
are used to derive statistical corrections to the data  
and are found to be relatively small,
typically $\lsim$ 5\% of the velocity dispersion  and 0.01~mag in the
\mg line-strength. 
Typical errors are about 8\% in velocity dispersion and
0.01~mag in \mg, in good agreement with values published elsewhere.

\end{abstract}


\section{Introduction}

If large-scale structures in the universe develop through
the action of gravity, their growth induces peculiar
velocities which are detectable as deviations of the galaxies' motion
relative to the smooth Hubble flow.  Therefore, by measuring redshifts
and redshift-independent distances for a large number of galaxies it is
possible to map the peculiar velocity field and to use it to probe the
characteristics of the underlying mass distribution 
as well as to constrain cosmological parameters by
comparing predicted and measured peculiar velocities (\eg Bertschinger \etal 
1990; Strauss \& Willick 1995; Nusser \& Davis 1994;
Willick \& Strauss 1998).

Following pioneering attempts 
(\eg Rubin \etal 1976; Tonry \& Davis
1981; Aaronson \etal 1982 ), the first successful measurement 
of peculiar motions
in the local universe was carried out by the ``7 Samurai''
(7S)  group, who developed the $D_n - \sigma$ distance method for
elliptical galaxies and showed 
in a series of papers (Dressler \etal 1987; Davies 
\etal 1987; Burstein \etal 1987; Lynden-Bell \etal 1988, Faber \etal 1989)
that the mass distribution in the local volume presents
significant velocity and mass density fluctuations.  The 7S sample is
an all-sky survey of about 400 early-type galaxies, generally brighter
than $m_B$=13.5 mag. Analysis of the measured velocity field led to the
discovery of the Great Attractor (GA) earlier conjectured by Tammann \&
Sandage (1985) and later shown to correspond to a large
concentration of galaxies in redshift space (da Costa \etal 1986, 1987;
Burstein, Faber, \& Dressler 1990;  Woudt \etal 1999).

These surprising results, led to the demise of the standard high-bias
CDM model, and questions raised by the 7S work motivated 
attempts to expand the samples of galaxies with measured distances, most
of which used Tully-Fisher (TF) distances to spirals 
(Willick 1990; Courteau \etal 1993). However, most of these more recent 
investigations either had limited sky coverage or used very sparse
samples.  Major progress only became possible after the completion of
wide-angle redshift-distance TF surveys such as those conducted by
Mathewson \etal (1992, 1996) and the SFI survey (\eg Haynes \etal
1999a, b) of spiral galaxies. These new surveys have been assembled to
produce homogeneous all-sky catalogs such as the Mark~III (Willick \etal
1997), the SFI catalog (\eg da Costa \etal 1996, Giovanelli \etal 1998),
and SHELLFLOW (Courteau \etal 2000).
These data have been extensively used in recent analyses, but 
in spite of the qualitative similarity of the recovered flow fields,
a quantitative comparison shows conflicting results,
illustrated by different estimates of the parameter
$\beta=\Omega^{0.6}/b$ where $\Omega$ is the cosmological density
parameter and $b$ is the linear bias factor relating galaxy and mass
density fluctuations  (\eg da Costa \etal 1998a, Zaroubi \etal 1997,
Willick \& Strauss 1998). Using the POTENT density-density method tends to 
produce higher values of $\beta$ than does the VELMOD velocity-velocity
technique although more recent results are more consistent 
(Zaroubi \etal 2002).

Recent work on the motions of early-type galaxies such as 
Lauer \& Postman (1994), M\"uller \etal (1998, 1999),
the EFAR survey (Wegner \etal 1996; Colless \etal 2001),  
and SMAC (Hudson \etal 1999) were designed to extend our knowledge
of the peculiar motions of galaxies to greater distances 
($R \sim 60h^{-1} - 110h^{-1}$Mpc, where 
$h \equiv H_0 / 100$ \kms Mpc$^{-1}$) in sparser
surveys some of which only cover part of the sky. For 
most of these investigations, the fundamental plane (FP) method was employed. 
More recent distance methods
such as Type Ia supernovae (Riess 2000) and surface brightness fluctuations
(Tonry \etal 2000) are more accurate than the methods related to TF or FP
on a per object basis, \eg $\Delta R/R \sim 5-7\%$ compared to $\sim20-25\%$
for $D_n-\sigma$, but have neither 
the numbers of objects nor the depth ($R \sim 50h^{-1}$ and $30h^{-1}$Mpc
respectively) to reconstruct the
velocity flows (see Dekel 2000 and Colless \etal 2001 for compilations
of recent peculiar motion surveys and their depths). Thus a new all-sky 
survey of early-type galaxies extending the earlier 7S sample has been needed. 

Such an all-sky survey has been the goal of the Redshift-distance
Survey of Nearby Early-type Galaxies (ENEAR). 
The sky coverage and selection of the survey have been
given in da Costa \etal 2000a (Paper I). It is an all-sky 
redshift-distance survey of
early-type galaxies, within $cz \leq 7000~$\kms, drawn from a $m_B=14.5$
magnitude-limited sample with complete redshift data. 
The completeness and selection criteria of the ENEAR sample have
been detailed in Paper I and given in Figure 10 of that paper which also
provides maps of sky coverage and other information on the survey and in
general, the sample completeness is nearly constant at $\ge$80\% for
the above magnitude and redshift limits.. 
During the course 
of the ENEAR project, it was necessary to add cluster
galaxies outside these criteria that were used to calibrate the 
$D_n - \sigma$ relation. These objects are more
distant and fainter than the criteria above used for the ENEAR selection
and have been described in Bernardi \etal 
(2002a,b). Further galaxies were also observed to compare our
data with the literature. 

With nearly three
times the number of galaxies and a fainter limiting magnitude, the ENEAR
survey has greater depth ($R \sim 60h^{-1}$Mpc) 
and resolution than the 7S study ($R \sim 30h^{-1}$Mpc) and it is
intended to complement the SFI spiral TF ($R \sim 65h^{-1}$Mpc)
survey with comparable depth and sky coverage but employing different
distance relations. Using early-type galaxies one hopes to
settle some of the pending issues. These include testing 
the universality of the
results from the SFI and related flow studies based on spiral galaxies
using a completely independent sample, based on a different
distance relation and a galaxy population which closely follows the
ridges of structures, in contrast to the more widely distributed spiral
population. 

A number of 
analyses have already been carried out using ENEAR data and indicate
that the cosmic flows found from the early-type and spiral galaxies
are indeed statistically equivalent. Borgani \etal (2000) studied the 
velocity correlation function of the ENEAR clusters and found they agree 
well with the SCI spiral cluster sample. Da Costa \etal (2000b) found that the 
dipoles defined by the flow of the ENEAR early-type 
galaxies agrees well with that of
the SFI spirals using TF distance measurements.
Nusser \etal (2001) compared ENEAR and IRAS PSCz velocity fields 
and obtained good agreement between the early-type and spiral
galaxy results. Zaroubi \etal (2001) studied the
large-scale power spectrum of the ENEAR velocity field and derived
the density field finding that most of the Local Group's motion is produced
by mass fluctuations within $80h^{-1}$~Mpc. 
Feldman \etal (2003) combined ENEAR data with other surveys to derive
$\Omega_m$ and $\sigma_8$.
Bernardi \etal (1998) studied the \mg line strengths of ENEAR
galaxies in three different density regimes, ranging from high to low, finding
that their galactic spheroids must have formed at redshifts ($z \gsim 3$)
independantly of their present environments. 
Bernardi \etal (2002a,b) described the construction of the $D_n - \sigma$
template from early-type galaxies in clusters which have been used to
estimate distances and derive peculiar velocities.  

Here we present the spectroscopic data of the ENEAR survey, which
complement the photometric data presented by Alonso et al. (2003). This paper
is organized as follows: Section 2 presents a brief description of the
sample, as well as the observations and reduction procedures used for
the spectroscopic data. Section 3 presents the techniques used to
measure redshifts, velocity dispersions and Mg2 indices; in this
section we also discuss the calculation of the internal errors, and
the corrections applied to the data due to observational effects. In
Section 4 we describe the procedure that was used to create the
master catalogue of homogeneous measurements with their estimated
errors. In Section 5 we present the calibrated and fully corrected
measurements. A brief summary follows in Section 6.

\section{The Spectroscopic Data}
\label{spec}

\subsection{ The Sample}

Here we present spectroscopic parameters for 1178 galaxies
from measurements obtained from 1701 spectra.  We have collected data for
a sample of early-type galaxies ($T\leq-2$ in the
Lauberts \& Valentjin 1989 system) which combines galaxies from: 1)
the ENEARm sample consisting of galaxies brighter than $m_B \leq
14.5$ within 7000~\kms (Paper I); 2) our measurements for galaxies in
cluster/groups in the ENEARc sample used by Bernardi \etal (2002a,b) 
that adhere to these same two selection criteria; 
3) galaxies in the SSRS2 (da Costa 
\etal 1998b) with adequate $S/N$ to reliably measure the velocity 
dispersion and line indices. Paper I and Bernardi \etal (2002a) also 
describe the ENEAR sample and subsamples. Figure 1 gives a histogram of the 
morphological types for the galaxies in this paper.

\subsection {Observations} \label{specobs}   

The spectroscopic observations reported here were made over several years
from three sites (CASLEO\footnote{Complejo
Astronomico El Leoncito (Argentina)},  ESO\footnote{European Southern
Observatory (Chile)}, and MDM\footnote{MDM 
Observatory (Arizona, USA)}).
During our program 30 spectroscopic observing runs
were carried out at the different sites. In total, 
there were 127 usable nights for spectroscopic observations in the period
1992-1999. Eleven setups were employed corresponding to different
combinations of telescope, detector, and spectrograph and the
resolution varied from about 
2 to 5 \AA. A total of 1701 spectra were obtained.
Table~\ref{tab:spectruns} summarizes the observations,
listing: in column (1) the run identification number; in column (2)
the date of the observations; in column (3) $N_s$ the number of 
spectroscopic nights for the run and in column (4) the setup reference
number described below.
  
Table~2
summarizes the different setups: in column (1) the setup
reference number; in column (2) the observatory and telescope used; in
columns (3) and (4) the number of spectra $N_s$ and the number $N_r$
of repeated observations using each setup; in columns (5)-(9) the
characteristics of the detector such as its identification, size,
pixel scale, gain and readout noise; in columns (10)-(13) the
characteristics of the spectrograph such as the slit width, the
grating, the dispersion, the resolution (as measured from the width of
the calibration lines), and the spectral coverage. All new spectra
were obtained using long-slits. The resolutions used can be divided into two
groups which we refer to as the high ($\sim 2.5$ \AA)
and low resolution ($\sim 5$ \AA), the low resolution
setups being 2, 3, 6, 7, and 10. Note that $\sim
80\%$ of the spectra were obtained at high resolution. 

As the goal of the new spectroscopic observations was to
measure both central velocity dispersions and line strengths, the spectral
range was chosen to cover the Mgb band (around
$\lambda_0 = 5177$ \AA), the E-band (5270 \AA), and the Fe I line
(5335 \AA). Most of our observations also included H$\beta$
(4861~\AA).

We followed standard procedures for observing the CCD spectra. 
In general identical observational and reduction procedures were 
used for ESO and CASLEO data. Wavelength
calibration lamps were observed before and after each object (He-Ar at
ESO; Hg-Ar-Xe-Ne at MDM; He-Ne-Ar at CASLEO). Dome flats, bias and dark
current frames  were taken nightly. The dark current was
checked for each CCD, but always found to be
negligible. Usually multiple exposures of a given galaxy were taken to
facilitate cosmic-ray removal. Exposure times varied with
object and sky conditions, but typical values were 10-20 minutes
at MDM and 20-30 minutes at ESO and CASLEO.

During each night stars with known radial velocities in the
spectral range of G8 to K5 and luminosity class III were observed 
for use as velocity templates. 
The MDM spectra were trailed along the slit and the ESO data consisted of
single exposures along the slit. We also observed a subset of the
Lick standards (Worthey \etal 1994) covering a wide range of spectral types.
Normally velocity standards were observed nightly and several were observed 
during each observing run. 

Determining velocity dispersions requires relatively high $S/N$ spectra
compared to that needed for redshifts alone.
Consequently for the galaxies we endeavored to obtain around 600 photons
per \AA, which corresponds to a signal-to-noise, $S/N$, 
$\sim$ 25, in continuum
bands near the Mg$_2$ features used for determining both $\sigma$ and
the \mg index. The quality of each observed spectrum was estimated from
the mean of the $S/N$ ratios measured at the
continuum bands 4895--4957 \AA\ and 5301--5366 \AA, in the vicinity of
the spectral region where the Mg$_2$ index is computed.  The resulting
distribution of $S/N$ for our sample in the \mg region is shown in
Figure~\ref{fig:SN_ENEAR}. The median $S/N$ of our
spectra is 26.8, slightly above our goal, with a rms scatter of 9.5, 
but with a large tail extending to higher $S/N$. 

Some of the brighter program galaxies were chosen as standards, and
were systematically observed at least once every run, when favorably
placed on the sky. This sub-sample contains $\sim$ 200 galaxies
that were observed twice or more
and the number of repeated observations range from 2 to 12 for a given galaxy.
These measurements were used to compare the low to high resolution
spectra and to place measurements obtained with different setups and
telescopes on a uniform internally consistent system. Figure~\ref{fig:histreob}
shows the distribution of repeated observations. 

\subsection{Data Reduction}

All spectra were reduced using the standard long-slit procedures in
the IRAF\footnote{IRAF is distributed by the National Optical
Astronomy Observatories which is operated by the Association of
Universities for Research in Astronomy, Inc. under contract with the
National Science Foundation} package. The reductions are
described briefly herein and follow standard methods (\eg
Wegner \etal 1999 where further details can be found) 
using the following steps: bias subtraction; flat field
correction; rejection of the cosmic-ray hits; wavelength calibration;
subtraction of the sky spectrum; and extraction of the one-dimensional
spectra. All but the last step was done on the two dimensional
images which provides line rectification. Each run was reduced by
one person and
even though similar, the reductions of the MDM and ESO/CASLEO
data were done independently with minor procedural differences, 
pointed out below.

Nightly sets of bias frames were scaled by the level of the CCD
overscan strip and medianed. These were checked for temporal
variations and then the resulting bias frame was subtracted from the
other images to remove the bias structure.  Because of the
stability of the systems at ESO, CASLEO,  and MDM, median bias frames 
could be constructed for the entire run and then subtracted from the
remaining frames.

Pixel-to-pixel sensitivity variations were removed by 
median filtering the flatfield exposures,
typically 10 or more per night. These were usually produced from
exposures of tungsten lamps either inside  the spectrograph 
as at MDM or an
illuminated target inside the dome at ESO and CASLEO, 
passing through the optics of the
spectrograph. A map was produced by normalizing the flatfield
constructed from the combined spectra relative to a smoothed version of
itself. The rms variation in the resulting flattened response frames was
typically less than $0.5\%$. Each galaxy or star spectral frame was then
divided by this response function map. 

Cosmic-ray hits were removed as follows. For MDM and most of the ESO
spectra the IRAF {\it lineclean} routine was employed. This
fits the galaxy's spectrum along the direction of the dispersion and
identifies cosmic-ray hits without affecting the absorption lines. For
CASLEO and some ESO spectra {\it imedit} was used to remove the cosmic ray
hits.

Wavelength calibrations were produced by fitting a
polynomial, typically 5th order, to the comparison spectra 
with a fitting accuracy of about $\pm$0.1
pixel. The wavelength calibrations generally employed more than 20
lines, and produced residuals of order $\pm0.02$ \AA\ for the ESO 1200
l/mm grating spectra which is representative for our observations.  The
wavelength calibration for ESO spectra used the set of He-Ar lines
compiled by M. P. Diaz available from
ftp://www.lna.br/pub/instrum/cass/hearlna.dat.Z. These gave consistently
better solutions than the standard tables distributed with IRAF.

Sky subtraction was facilitated using the sky level determined in the
IRAF {\it background} routine from two or more regions on each side of
the galaxy spectrum far enough not to be contaminated by the
object itself. The sky level at the object was interpolated using a
low order polynomial fitted to the sky in a direction perpendicular to
the spectrum.

Each final one-dimensional galaxy spectrum was then extracted by summing
across its profile on the CCD image in the region where it was greater
than about 5\% of its maximum using the IRAF task {\it apsum}
with  the variance weighting option. 
For some of the ESO observations the object spectra were
extracted by summing the  region where the galaxy
flux approaches the sky level and
the sky value was determined from the median value measured
in two regions on each side of the galaxy, and then interpolated
across the galaxy spectrum.

Finally, all one-dimensional spectra were visually inspected. About
10\% of the observed galaxies show some  emission lines
characteristic of HII regions, while others exhibit features typical 
of A and F stars (hydrogen Balmer-line absorption). These cases are  
listed in the comments to Table 4. 

\section {Spectroscopic Parameters}
\label{parameters}

\subsection{Redshifts and Velocity Dispersions}

The measurements of the redshift, $cz$, and the velocity dispersion,
$\sigma$, were obtained using the IRAF task {\it fxcor} in the {\it
RV} package. This task employs the Tonry \& Davis (1979)
cross-correlation technique which generally yields more robust
measures for modest $S/N$ spectra than other more complicated algorithms
such as the Fourier coefficient (\eg Rite\'~1999). Each spectrum is
linearized in $\log{\lambda}$, has the continuum removed by a
low-order polynomial, and is end-masked with a cosine bell function
prior to the cross-correlation analysis. Following Baggley (1996) and
Wegner \etal (1999) the measurements of redshift and velocity
dispersion are carried out in two steps.  A first estimate of the
redshift and FWHM is 
obtained using the whole observed spectrum. Next, using the first
redshift estimate an improved measurement of the redshift and of
the FWHM of the cross-correlation peak is obtained by
restricting the wavelength range. For each galaxy-template combination the
FWHM of the correlation peak is calculated using the spectral region
with rest wavelength 4770--5770 \AA. This FWHM is then calibrated by
convolving each standard star's spectrum with a series of Gaussian
broadening functions to construct a curve
relating the cross-correlation peak FWHM with the input $\sigma$ value.

Internal errors in the measurement of the velocity dispersions 
arise from systematic
errors associated with the template-galaxy mismatches and the
statistical errors due to the noise properties of the
spectra. The errors in the $\sigma$'s were estimated 
by calibrating the Tonry \& Davis (1979) $R$ value, the height of the 
true peak to the average peak in the cross-correlation, using
simulated spectra with different noise values and indicates that our error
estimates depend on
$S/N$ and the velocity dispersion. The velocity dispersion 
dependence arises because at low-$\sigma$ one is limited by the
instrumental resolution, while at high-$\sigma$ the absorption lines
broaden leaving only a small contrast relative to the
continuum. Both effects tend to increase the amplitude of the
error.

The internally defined error is normalized
on a run by run basis from the ratio of the standard deviation of
repeated exposures of the same galaxy, observed in the same run and
with approximately the same $S/N$, to the internal error estimate. All
internal errors for that run are multiplied by this
factor. Figure~\ref{fig:sigerr} shows our final estimates of the
fractional error $\delta \sigma/\sigma$ as a function of $\sigma$ the velocity
dispersion (left panel) and the $\delta\sigma / \sigma$ distribution
(right panel), for all the observed galaxies. As can be seen on the left side
for $\log \sigma \gsim ~2.2$~ the errors are essentially constant but then rise
at the low-$\sigma$ end which comprises less than 10\% 
of the sample. 

\subsection{Aperture Corrections}
The velocity dispersions were corrected by applying an
aperture correction to the observed velocity dispersion. 
This accounts for the dependence of the measured
velocity dispersions on: 1) observational parameters such as
the seeing and the size and shape of the spectrograph slit; 2) the
galaxy's distance, since a fixed slit size projects to 
different physical scales on galaxies with distances;
3) the intrinsic velocity and luminosity profiles
of the galaxy. Expressions for the aperture correction were
obtained empirically by Davies \etal (1987) and by J{\o}rgensen \etal
(1995b) using kinematical models. Here we adopt the latter's metric
aperture correction:
\begin{equation}
\label{eq:apertjfk}
\log \left( \frac{\sigma_{cor}}{\sigma_{obs}} \right) = 0.038 \log
\left[ \left( \frac{r_{ap}}{r_{norm}}\right) \left(
\frac{cz}{cz_o} \right) \right] 
\end{equation}
where $\sigma_{obs}$ is the value of the velocity dispersion observed
through an equivalent circular aperture of $r_{ap}$, which 
for a rectangular slit is 
$r_{ap}=1.025\sqrt {wl/\pi}$ in arcsec, $w$ and $l$ being the
width and length of the slit and $\sigma_{cor}$ is the corrected value
normalized to a circular aperture of radius
$r_{norm}=0.595~h^{-1}$~kpc,
$cz$ is the redshift of the galaxy, and $cz_o$ is a reference redshift
taken to be that of Coma ($cz_0 = 7010$ \kms). The standard aperture
corresponds to 1.7 arcsec at the Coma distance.  

\subsection{Line Strengths}

We have also measured the Mg$_2$ index and scaled it to the Lick system, for
all the available spectra. This line index is an
indicator of metallicity and star-formation rate (\eg Bernardi \etal
1998; Colless \etal 1999).  
The Mg$_2$ index is given in
magnitudes and measures the depression of the spectral intensity 
due to the combined broad Mg~H feature and the Mgb triplet and is 
defined as
\begin{equation}
{\rm Mg}_2 = -2.5 \log _{10} \frac{\int_{\lambda
_{1}}^{\lambda _{2}}S(\lambda)/ C(\lambda) d\lambda}{\Delta \lambda}
\end{equation}
where $\Delta \lambda =\lambda _{2}-\lambda _{1}=42.5 $~\AA is the
width of the Mg$_2$ bandpass ($5154.1-5196.6$ \AA), $S(\lambda)$ is
the object spectrum and $C(\lambda)$ is a
pseudo-continuum. Following Gonz\'ales (1993) and
Worthey \etal (1994) the pseudo-continuum is estimated by a linear
interpolation between the mid-points of the side bands
($4895.1-4957.6$ \AA ~and $5301.1-5366.1$ \AA) where
the average flux is computed within these two side bands.

Most of our spectra lacked spectrophotometric flux calibrations and had
resolutions higher than the
Lick system, so we adopted the following procedure to measure the \mg
line index.  First, all spectra were degraded in resolution by smoothing with
a Gaussian filter with a width chosen to match the  
spectral resolution of the Lick/IDS (8.6 \AA). Second
the detector response was accounted for on a
run by run basis.  For each run a 
low-order polynomial (1-3) was fit to
the spectra of galaxies in common with Faber \etal (1989)
over a wavelength range of about 500 \AA. The order of the fit was
chosen so that after dividing the observed spectra by this polynomial,
and measuring the \mg index, a good
agreement with Faber \etal (1989) was
obtained. This polynomial was then retained for all spectra in the run, 
leaving the zero-point free. Figure 5 shows the resulting 
differences between our measured \mg line
indices and those of Faber \etal (1989).
The order of the polynomial
depends on the resolution. For our low-resolution
spectra a linear fit worked well, while a polynomial 
of order greater than 3 was
required in the case of high resolution spectra.

In the \mg indices
we find no significant zero-point shift and a relatively
small scatter of 0.015~mag.  We have also measured the \mg index
directly, ignoring possible variations in the response function
for runs with available Lick
standards.  In these cases the line index is computed for the stars,
the resulting value is then corrected to the Lick values and the same
correction applied to the galaxies. The two methods
lead to consistent results with a scatter of about 0.014~mag,
comparable to that obtained from the comparison with galaxies
measured in the Lick system by Faber \etal (1989).

The \mg line strength errors were estimated using simulated spectra.
For each run all high $S/N$ stellar templates
were used to generate a set of spectra of different $S/N$ and velocity
dispersions.  This was done by adding Poisson noise and convolving with
Gaussians of varying width to simulate galaxies with different
velocity dispersions. For each template a total of about 1000
simulated spectra were generated in 50~\kms intervals of velocity
dispersion and $S/N$ ranging from 10 to 60. For each template,
$\sigma$ and $S/N$ the rms value of the measurement of the \mg index,
following the same procedure adopted above, was computed. Thus an error grid
was generated for each template. The error in the \mg
measurement for an object was taken to be the largest value at the
appropriate value of $\sigma$ and $S/N$. 

Figure 6
shows resulting distribution of the estimated errors, $\delta$Mg$_2$, 
in the
measurement of the Mg$_2$ line index for all of our galaxies 
found using the procedure described above. We find
that the median error is $ 0.013\pm 0.002$~mag, comparable to the
values obtained by other authors (\eg 
J{\o}rgensen, Franx, \& Kj{\ae}rgaard 1995a, b; Colless \etal 1999).

Our final \mg line indices are corrected for aperture
effects and for the broadening of the line due to the velocity
dispersion of the galaxies, which underestimates the value of the
index for high-$\sigma$ galaxies (\eg Gonz\'ales 1993). Following 
J{\o}rgensen, Franx, \& Kj{\ae}rgaard 1995b) we have adopted an aperture
correction for the \mg index which is similar to that used for 
the velocity dispersion:
\begin{equation}
 {\rm Mg}_2^{cor}-
{\rm Mg}_2^{obs}=  0.038 \log
\left[ \left( \frac{r_{ap}}{r_{norm}}\right) \left(
\frac{cz}{cz_o} \right) \right] 
\end{equation}

The $\sigma$ broadening correction used here was derived as
follows. The spectra of standard stars available in a run were first
convolved with Gaussians of different dispersions. Next the ratio
between the value of the index as measured in the original
un-convolved spectra to that measured on the convolved spectra was
determined as function of the velocity dispersion. A smooth curve was fitted to
the ratios obtained for different templates. The correction for a
galaxy of a given $\sigma$ was obtained from this fit.  All
runs have shown a similar correction of $\sim 0.001$~mag at
$\sigma=100$~\kms, which increases approximately linearly to
$\sim~0.004$ mag at $\sigma=400$~\kms.

\section{Internal and External Comparisons}
\label{comp}

In order to make our spectroscopic measurements 
from different runs internally consistent, we find that 
only relative zero-point shifts are necessary. 
Therefore, measurements obtained in
different observing runs are brought onto a common system by applying
these zero-point corrections. This procedure takes into  
account the number of overlaps available at each site and for each setup.
It optimizes the number of overlaps in the comparison to improve
the statistics in the determination of the offset required to bring them
into a common system. The high-resolution ESO data (setup~4
in Table~2) are taken as the reference. These data were
chosen as the fiducial system because they have the highest resolution,
comprise the largest number of spectra in our sample, and have the
greatest number of galaxies with repeated observations in common with
other instrumental setups.

To determine the ``fiducial'' system we corrected our spectroscopic
parameters using the mean difference $\Delta x_i$ of the measurements of
run $i$ with all the other runs $j \ne i$ for galaxies in run $j$ in common
with those in $i$. This offset is computed with variance weighting using
the estimated errors in each measurement:
\begin{equation}
\Delta x_i = \epsilon_i^2\sum_{j \ne i} \sum_{k \epsilon i,j} 
\frac {x_{i,k}-x_{j,k}} {\Delta x_{i,k}^2+\Delta x_{j,k}^2}
\end{equation}
Here $k$ runs over the galaxies in common between runs $i$ and $j$, and
$x_{i,k}$ corresponds to the measurement of either $\log \sigma$ or \mg
for galaxy $k$ in run $i$ and $\epsilon_{i}$ is the 
standard error in the mean, estimated by:
\begin{equation}
\epsilon_i = \left (\sum_{j \ne i} \sum_{k \epsilon i,j} 
\frac{1}{\Delta x_{i,k}^2+\Delta x_{j,k}^2}
\right )^{-\frac{1}{2}}
\end{equation}

We determine the most significant offset by finding the run with the
maximum value of $\Delta x_i/ \epsilon_i$, and iterate towards a common
zero-point by subtracting this offset from the measurements of run $i$. 
We halted the process when the most significant offset was $\Delta x_i/
\epsilon_i <2$. About four iterations were required
for redshift, velocity dispersion, and the Mg$_2$ index
parameters. These corrections are relatively small
amounting to $<  25$ \kms with a scatter of 40 \kms for redshift,
$\lsim 0.025$ dex with a scatter $\lsim 0.060$ dex for $\log \sigma$, and 
$\lsim 0.015$ mag with a scatter of $\sim0.020$ mag for $\rm{Mg}_2$.  
After defining the fiducial system, we compare aperture corrected
values, whenever necessary, obtained using different setups at ESO. 

The relatively large number of overlapping observations provide the
necessary information to derive suitable statistical corrections for all
runs at different sites.
When making this comparison, we used the convention of performing the 
differences between ``older-newer'' measurements. For instance, we
compared a measurement taken in the first ESO-651 run with all the 
subsequent ESO runs. We then compared the second run
(ESO-652) measurements with all later runs
(ESO-653, 654, etc. but not ESO-651)and so on. The
measurements obtained from MDM and CASLEO spectra were corrected as
follows: for runs with a significant number of galaxies in common
with our reference system, the measured values were directly compared to
this system, while for other runs where the number of galaxies in
common is small, the comparison was made using calibrated
measurements for that telescope and setup.  

The offsets derived from the comparison of all other runs not used in
the definition of the fiducial system are small and therefore 
consistent with those offsets found in defining the reference system.
This indicates the high degree of homogeneity of the data. Only one 
CASLEO run, which contributes the least to the overall sample, 
required a large offset correction, $\Delta \log \sigma = 0.064$ dex.  

The final results of the uniformization are
presented in Figure~\ref{fig:comparsp_int} which shows the comparison
between the ESO measurements of $\log \sigma$ (left panels) and  
\mg line index (right panels) with those obtained from
ESO, MDM, and CASLEO spectra (from top to bottom). 
The results are summarized in Table~3 which gives: in
column (1-2) the sites; in column (3) the number of repeated measurements
$N_m$ in the same or in different runs; in columns (4) and (5) the mean
offset and its error of the differences of the calibrated $\log \sigma$
and (6) and (7) of the \mg measurements. 
These results show that the corrections lead to an
internally consistent system with only a small ($\lsim$ 1\%) residual
offset in the velocity dispersion. The last two rows report
the internal comparisons for MDM and CASLEO. 

After bringing all measurements to a consistent system, multiple
measurements of the same galaxy are combined weighting by their individual
errors as described in the next section. These final values
are then compared with those of previous studies in the literature as
presented in Figures 4 and 5 of Bernardi \etal (2002a). 
As shown in that paper, we find an overall residual difference 
between ENEAR measurements and those in the literature of -0.002$\pm 0.004$~dex
and a scatter of 0.051 in $\log (\sigma)$. For \mg we find an offset of
$0.003\pm0.002$~mag and a scatter of 0.018~mag.  These observed
scatters are consistent with an error per galaxy of about 8\% in
velocity dipersion and 0.01~mag in \mg. Note, however, that most of the
available data in the literature is limited to values $\gsim~100$~\kms
and \mg $\gsim 0.18$~mag. As seen in the 
internal estimates above, we expect increasing measurement errors for
these smaller quantities as one approaches the resolution
limit.

\section{The Spectroscopic Catalog}

The final value of each of the spectroscopic parameters for a galaxy with
multiple observations is given by the error weighted mean of the
individual measurements. The error for these galaxies is 
computed by adding in quadrature the error associated to the mean
with the rms scatter of the repeated measurements. Whenever necessary,
values which differ by more than three times the $rms$ from the mean
were removed to avoid biasing the results due to a few outliers.
For small values of $\sigma$ and \mg only the measurements obtained at
high-resolution are used.

Table 4 lists the final fully corrected and, if more than
one observation is available, combined spectroscopic data for 1178
galaxies from the ENEAR observations; no literature data are used
\footnote{In comparing data in Table 4 with those in
common with Bernardi \etal (2002b), some differences occur due 
to two causes. Firstly Bernardi \etal include literature data in the 
averages and in this paper only ENEAR data are included. Secondly Table 4
is a later compilation of ENEAR spectroscopic data. Some
additional observations were added and the homogenization of the runs to a 
common system were recomputed; differences were usually small or unchanged.
}.
The photometric portion of the ENEAR survey is given in Alonso \etal (2003);
it should be noted however that not all objects have both kinds of data.
The table shows: in column (1) and the galaxy standard
name; in columns (2) and (3) the (2000) equatorial
coordinates; in column (4) the morphological parameter $T$ (see 
Paper I); in column (5) the photographic magnitude $m_{B}$; in column (6) 
$N_{obs}$,the number of spectra used for redshift and $\sigma$; 
in columns (7) and (8) the heliocentric redshift and error;
in columns (9) and (10) the velocity dispersion and error; in
column (11) $N_{Mg_2}$ the number of spectra used to determine 
the \mg line index;
in columns (12) and (13) the Mg$_2$ line index and its error; in column
(14) notes; and column (15) denotes whenever the galaxy has data 
form the literature.
Here we present only the first few
entries of the table which can be retrieved from the electronic version
of this journal.

The redshift, velocity dispersion and \mg line
index distributions  for the 1178  galaxies listed in Table 4 are shown
in Figure~\ref{fig:hist_sigmg2}. The sharp break seen in the redshift
distribution at $cz = 7000$~\kms reflects the redshift cutoff 
of the ENEARm sample. Galaxies beyond this redshift are in
clusters or are fainter than $m_B = 14.5$. Also note
that a significant number of galaxies ($\gsim 100$) have been measured
at the low-$\sigma$ ($\lsim 100$ \kms) and small line index end ($\lsim$
0.20) where the number of such galaxies with measured values 
in the literature is remarkably small.

The individual measurements used to construct Table~4 are given in Table~5,
for which we also present the first few entries and the entire table can be 
obtained from the electronic version of this journal. These measurements  
include the run corrections described above and the table 
contains: column (1) is the galaxy standard name; columns (2) and (3) give the
(2000) equatorial coordinates; column (4) is the magnitude m$_{B}$; 
column (5) is the morphological type (T);
column (6) is the run number from Table 1; columns (7) and (9) contain the 
heliocentric redshift and error; columns (9) and (10) are $\log \sigma$
and error; and columns (11) and (12) are the measured \mg line index
and error.

\section{Summary}

\label{summary}
We have presented  spectroscopic observations for the ENEAR project
and described their reduction and quality assessment. There are 1701
spectra of 1178 galaxies, of which $\sim80$\% had no previous
measurements of redshift, velocity dispersion, and \mg line index. 
In addition to the velocity dispersions, we have measured the \mg
index for 1149 galaxies. About 80\% of the observations were conducted
with a resolution ($\lsim $2.5\AA) which is a factor of 2 better 
than previous large
surveys. The observations span a number of years utilizing different
instruments, but repeated
observations allow the measurements to be brought into a
common system that is internally consistent and compares well with
published data. From the comparison with external data we confirm
our error estimates  which are typically of $\sim 8\%$ in $\sigma$ and
0.01~mag in \mg. The errors are nearly constant for $\sigma > 100$~\kms
and \mg$>$ 0.2~mag, increasing for smaller values.
 
Since there is considerable overlap with measurements of velocity
dispersion and \mg by other authors  (Bernardi \etal
2002a,b), it is possible to derive statistical corrections which can be
applied to these other  measurements to produce a uniform catalog of
about 2000 early-type galaxies with measured velocity dispersions and
1300 with measured \mg line index. Such a sample is an invaluable
database for studies of the properties of the early-type galaxies and
their peculiar motions. Our sample is currently one of the largest
uniform data sets of spectroscopic measurements of nearby early-type
galaxies.

GW acknowledges support from the following over the
course of this project: Dartmouth College, 
the Alexander von Humboldt-Stiftung for a year's
stay at the Ruhr Universit\"at in Bochum, and ESO for supporting trips to
Garching.
MB thanks the Sternwarte M\"unchen, the Technische
Universit\"at M\"unchen, ESO Studentship program, and MPA Garching for
their financial support during different phases of this research.
CNAW acknowledges
partial support from CNPq grants 301364/86-9, 453488/96-0, and NSF
AST-9529028 and NSF AST-0071198.
MVA would like to thank the hospitality of the Harvard--Smithsonian 
Center for Astrophysics, the ESO visitor program and ON. 
We wish to thank the support of the CNPq--NSF bilateral program
(MVA, LNdC), a research fellowship from CNPq (PSSP) and CLAF (MVA, PSSP and
MAGM) for financial support to the project. MVA also acknowledges financial
support from the SECYT and CONICET (Argentina). Nearly all of the southern
observations were carried out using the 1.54 m ESO telescope thanks to an
agreement between ESO and the Observat\'orio Nacional.

{}

\newpage

\normalsize
\begin{table}
\begin{center}
\caption{Observing runs for spectroscopy}
\vspace*{10truemm}
\small
\begin{tabular}{lcrr}
\hline \hline
\\[-6mm]
\label{tab:spectruns}
Run & Date & N$_{s}$ & Setup\\
 (1) &  (2) & (3) & (4) \\
\\[-6mm]
\hline
\\[-6mm]
\\
MDM-501 & 1992 Oct      &5&  5\\
\\[-10mm]
ESO-651 & 1993 Nov      &6&  1\\
\\[-10mm]
ESO-652 & 1994 May      &7&  1\\
\\[-10mm]
MDM-502 & 1994 Oct      &6&  6\\
\\[-10mm]
ESO-654 & 1995 May      &1&  2\\
\\[-10mm]
ESO-653 & 1995 Aug      &4&  1\\
\\[-10mm]
MDM-503 & 1995 Dec      &4&  7\\
\\[-10mm]
CASLEO-801 & 1996 Apr  &3& 11\\ 
\\[-10mm]
CASLEO-802 & 1996 Sep  &3& 11\\
\\[-10mm]
ESO-655 & 1996 Oct     &5&  4\\
\\[-10mm]
ESO-656 & 1996 Nov      &13&  3\\
\\[-10mm]
MDM-505 & 1996 Nov      &3&  8\\
\\[-10mm]
ESO-657 & 1997 Jan      &5&  3\\
\\[-10mm]
MDM-506 & 1997 Feb      &2&  9\\
\\[-10mm]
ESO-658 & 1997 Mar      &6&  4\\
\\[-10mm]
ESO-659 & 1997 Apr     &10&  4\\
\\[-10mm]
CASLEO-803 & 1997 May   &5& 11\\ 
\\[-10mm]
MDM-507 & 1997 Jun    &5& 10\\
\\[-10mm]
ESO-660 & 1997 Oct      &5&  4\\
\\[-10mm]
MDM-508 & 1997 Nov      &3&  9\\
\\[-10mm]
ESO-661 & 1998 Feb      &6&  4\\
\\[-10mm]
ESO-662 & 1998 Apr      &7&  4\\
\\[-10mm]
MDM-509 & 1998 Apr/May  &2&  9\\
\\[-10mm]
ESO-663 & 1998 Jun      &3&  4\\
\\[-10mm]
ESO-664 & 1998 Aug      &2&  4\\
\\[-10mm]
ESO-665 & 1998 Oct      &4&  4\\
\\[-10mm]
MDM-510 & 1998 Nov      &1&  9\\
\\[-10mm]
ESO-666 & 1999 Feb      &11&  4\\
\\[-10mm]
ESO-667 & 1999 Aug      &2&  4\\
\\
\hline
\\
\end{tabular}
\end{center}
\footnotesize
Notes: 
Column (3) shows the number of spectroscopic nights for the run. 
Information about the setup indicated in column (4) is given in 
Table~2.
\end{table}

\newpage

\begin{deluxetable}{ccrrllcccccccc}
\tablecolumns{14}
\rotate
\tablewidth{0pt}
\tablenum{2}
\tabletypesize{\tiny}
\tablecaption{Observing Setups}
\tablehead{
Setup &Telescope & N$_{m}$ & N$_{r}$ &  Detector & Size & Spatial scale & Gain
& Readout Noise & Slit width & Grating & Dispersion & Cover Resolution & Spectra
l Range\\
                &          &         &         &           &      &
arcsec/pixel  & e$^-$/ADU & e$^-$ & arcsec & $l/$mm   & \AA/pixel  &
\AA          &   \AA \\
  (1) &  (2) & (3) & (4) & (5) & (6) & (7) & (8) & (9) & (10) &  (11)
  & (12)& (13) & (14)}
\startdata
1 & ESO 1.52    & 237 & 65 & CCD \#24   & 2048$\times$ 2048
&0.72 &2.9 & 8 &2.5 &600 &0.93 & 2.33 & 4500--6300\\
2 & ESO 1.52    &  4 & - & CCD \#24   & 2048$\times$2048
&0.72 &2.9 &8 &2.5 &600 & 1.87 & 4.68 & 3750--7300 \\
3 & ESO 1.52    & 114 & 10 &  CCD \#39   & 2048$\times$2048 
&0.72 &1.2 &5.5 &2.3 &600 &1.91 & 4.97 & 3750--7300\\
4 & ESO 1.52    & 831 & 218 & CCD \#39  & 2048$\times$2048
&0.72 &1.2 &5.5 &2.5 & 1200 &0.98 & 1.90 & 4300--6200\\
\\
  &        & 1186 & 293 &         &       &   & & & & & & & \\ 
\\
5 & MDM 2.4     & 30 & 6 &  Willbur   & 2048$\times$ 2048
&0.172&1.94&4.73 & 1.7 &600 &0.99 & 2.00 & 5180--7200 \\
6 & MDM 2.4    & 47 & 4 &  Willbur   & 1024$\times$ 1024
&0.343 &2.43 &4.73 & 1.7 &600 &2.81 & 5.60 & 4200--7000 \\
7 & MDM 2.4    & 56 & 7 &  Charlotte  & 1024$\times$1024 
&0.28 & 3.16&5.45 & 1.7 &600 &2.24 & 4.48 & 4500--6800 \\
8 & MDM 2.4    & 53 & 4 &  Templeton  & 1024$\times$1024 
& 0.28 & 3.47 & 5.33 & 1.7 & 1200 & 1.00 & 2.50 &4800--5800\\
9 & MDM 2.4    & 173 & 48 &  Charlotte & 1024$\times$1024 
& 0.28 & 3.16 & 5.45 & 1.7 & 1200 & 1.00 & 2.50 &4800--5800\\ 
10 & MDM 1.3   & 50 & 4 &  Charlotte & 1024$\times$1024 
& 0.51 & 3.16 & 5.45 & 1.2 & 600 & 2.10 & 4.50 & 4358--6882\\
\\
  &        & 409 & 73 &          &       &   & & & & & & & \\ 
\\
11 & CASLEO  2.15& 106 & 13 & Tek     & 1024$\times$1024
& & 1.98 & 7.4 &3 & 600 &1.62 & 3.41 &4500--6100\\
\\
  &        & 106 & 13 &          &       &   & & & & & & & \\
\\
Total      &    & 1701 & 379 &          &       &            \\ 
\enddata
\tablecomments{Run 6 used 2X2 pixel binning. Runs 5, 6, 7, and 10
used the Mark III spectrograph. Runs 8 and 9 used the Modspec spectrograph.
}
\end{deluxetable}

\newpage

\begin{planotable}{llrcccc}
\tablewidth{0pt}
\tablenum{3}
\label{tab:scomp}
\tablecaption{Internal Comparisons}
\scriptsize
\tablehead{Site 1 & Site 2 & $N_{m}$ & $< \Delta log{\sigma} >$ & rms/$\sqrt(2)$ &
  $< \Delta$ Mg$_2 >$ & rms$/\sqrt(2)$\nl}
\startdata
 ESO & ESO & 745 & 0.004$\pm 0.002$ & 0.038 & -0.001$\pm 0.001$ & 0.018\nl
 ESO & MDM & 110 & -0.006$\pm 0.006$ & 0.038 & -0.001$\pm 0.002$ & 0.020\nl
 ESO & CASLEO & 89 & 0.008$\pm 0.006$ & 0.035 & 0.003$\pm0.003$& 0.019 \nl
\nl
 MDM & MDM & 77 & 0.002$\pm 0.005$ & 0.029 & 0.001$\pm 0.003$ & 0.019\nl
 CASLEO & CASLEO & 13 & 0.011$\pm 0.013$ &  0.033 & -0.001$\pm 0.006$ & 0.016\nl
\enddata
\end{planotable}

\begin{deluxetable}{lccrrcrcccccccc}
\tablecolumns{15}
\rotate
\tablewidth{0pt}
\label{tab:data}
\tablenum{4}
\tabletypesize{\scriptsize}
\tablecaption{The Spectroscopic ENEAR Catalog}
\tablehead{
Name &$\alpha$ &$\delta$&                               T &  m$_{B}$& N$_{obs}$ & $cz_{hel}$ &                            $\epsilon_{cz_{hel}}$  & $\log_{10} \sigma$ &                  $\epsilon_{\log_{10} \sigma}$& N$_{Mg_2}$ & Mg$_2$ &           $\epsilon_{\rm{Mg}_2}$& Notes & Lit\\
 & (2000) & (2000) & & mag &  & $km s^{-1}$ &             $km s^{-1}$ &$km s^{-1}$ &$km s^{-1}$ & &  & & & \\
\tablevspace{5pt}
(1) & (2) & (3) & (4) & (5) & (6) & (7) & (8)             & (9) & (10) & (11) &                                             (12) & (13) & (14)  & (15)}
\startdata
  ESO409G012    & 00:04:42.2 & -30:29:00& -5 & 14.23 &  1 &  8044 &  63 &  2.386 & 0.040 &  1 &  0.242 & 0.011 & 1 &   *  \\      
IC1529        & 00:05:13.3 & -11:30:12& -2 & 14.50 &  3 &  6726 &  25 &  2.258 & 0.022 &  3 &  0.254 & 0.012 & 1 &   \, \\      
NGC7832       & 00:06:28.4 & -03:42:58& -3 & 13.50 &  2 &  6202 &  19 &  2.351 & 0.032 &  2 &  0.285 & 0.013 & 1 &   \, \\      
UGC00061      & 00:07:23.8 & +47:02:26& -2 & 14.30 &  2 &  5354 &  52 &  2.312 & 0.031 &  2 &  0.303 & 0.011 & 1 &   \, \\      
NGC0043       & 00:13:00.8 & +30:54:55& -2 & 13.90 &  2 &  4846 &  20 &  2.298 & 0.042 &  2 &  0.323 & 0.010 & 1 &   \, \\      
UGC00130      & 00:13:56.9 & +30:52:58& -7 & 14.20 &  1 &  4792 &  30 &  2.146 & 0.042 &  1 &  0.273 & 0.012 & 1 &   \, \\      
NGC0050       & 00:14:44.5 & -07:20:38& -3 & 12.50 &  1 &  5468 &  22 &  2.422 & 0.026 &  0 &  0.000 & 0.000 & 1 &   \, \\      
NGC0063       & 00:17:45.6 & +11:27:01& -5 & 12.60 &  2 &  1167 &  26 &  1.875 & 0.060 &  2 &  0.111 & 0.024 & 4, 3 &   \, \\   
NGC0068       & 00:18:18.7 & +30:04:17& -3 & 14.05 &  1 &  5790 &  29 &  2.414 & 0.032 &  1 &  0.304 & 0.009 & 1 &   \, \\      
NGC0078A      & 00:20:25.8 & +00:49:34& -2 & 14.50 &  1 &  5454 &  26 &  2.398 & 0.033 &  1 &  0.308 & 0.009 & 1 &   \, \\      
NGC0108       & 00:25:59.0 & +29:12:41& -2 & 13.30 &  2 &  4776 &  25 &  2.197 & 0.032 &  2 &  0.264 & 0.013 & 1 &   \, \\      
NGC0113       & 00:26:54.5 & -02:30:03& -3 & 14.00 &  1 &  4372 &  25 &  2.161 & 0.036 &  0 &  0.000 & 0.000 & 1 &   \, \\      
NGC0125       & 00:28:50.1 & +02:50:17& -2 & 13.83 &  1 &  5263 &  24 &  2.105 & 0.062 &  1 &  0.212 & 0.008 & 1 &   \, \\      
NGC0128       & 00:29:15.0 & +02:51:51& -2 & 12.92 &  1 &  4210 &  21 &  2.383 & 0.029 &  0 &  0.000 & 0.000 & 1 &   \, \\ 
\enddata
\tablecomments{(1)  no problems;
(2) star along the slit;
(3) emission lines;
(4) low S/N measurements;
(5) low velocity dispersion on the limit of the resolution;
(6) old data (Reticon);
(7) peculiar spectrum: eg. broad lines (supernova?),
absorption lines too weak or undetectable.
Table 4 is presented in its entirety in the electronic edition of the 
Astronomical Journal. A portion is shown here for guidence regarding its 
form and content.
}
\end{deluxetable}

\begin{deluxetable}{lccrrcrccccc}
\tablecolumns{12}
\rotate
\tablewidth{0pt}
\tablenum{5}
\tabletypesize{\scriptsize}
\tablecaption{Individual ENEAR Spectroscopic       Measurements}
\tablehead{
Name &$\alpha$ &$\delta$&                               T &  m$_{B}$& Run & $cz_{hel}$ &                                  $\epsilon_{cz_{hel}}$  & $\log_{10} \sigma$ &                  $\epsilon_{\log_{10}\sigma}$& Mg$_2$ &                         $\epsilon_{\rm{Mg}_2}$ \\ 
 & (2000) & (2000) & & mag &  & $km s^{-1}$ &             $km s^{-1}$ &$km s^{-1}$ &$km s^{-1}$ & mag & mag \\ 
\tablevspace{5pt}
(1) & (2) & (3) & (4) & (5) & (6) & (7) & (8)             & (9) & (10) & (11) &                                             (12) }
\startdata
ESO409G012    & 00:04:42.2 & -30:29:00               & -5 & 14.23 & 660 &  8044 &  63 &  2.386 & 0.040 &  0.242 & 0.012  \\
IC1529        & 00:05:13.3 & -11:30:12               & -2 & 14.50 & 653 &  6735 &  30 &  2.251 & 0.029 &  0.257 & 0.011  \\
              &            &                         &    &       & 656 &  6725 &  22 &  2.259 & 0.018 &  0.248 & 0.006  \\
              &            &                         &    &       & 653 &  6724 &  23 &  2.260 & 0.016 &  0.270 & 0.011  \\
NGC7832       & 00:06:28.4 & -03:42:58               & -3 & 13.50 & 653 &  6202 &  22 &  2.368 & 0.034 &  0.276 & 0.013  \\
              &            &                         &    &       & 653 &  6203 &  16 &  2.343 & 0.024 &  0.290 & 0.009  \\
UGC00061      & 00:07:23.8 & +47:02:26               & -2 & 14.30 & 508 &  5362 &  57 &  2.311 & 0.036 &  0.296 & 0.013  \\
              &            &                         &    &       & 508 &  5350 &  47 &  2.312 & 0.024 &  0.304 & 0.004  \\
NGC0043       & 00:13:00.8 & +30:54:55               & -2 & 13.90 & 508 &  4854 &  22 &  2.287 & 0.051 &  0.317 & 0.011  \\
              &            &                         &    &       & 508 &  4843 &  16 &  2.301 & 0.027 &  0.325 & 0.006  \\
UGC00130      & 00:13:56.9 & +30:52:58               & -7 & 14.20 & 505 &  4792 &  30 &  2.146 & 0.042 &  0.273 & 0.014  \\
NGC0050       & 00:14:44.5 & -07:20:38               & -3 & 12.50 & 501 &  5468 &  22 &  2.422 & 0.026 &  0.000 & 0.004  \\
NGC0063       & 00:17:45.6 & +11:27:01               & -5 & 12.60 & 651 &  1143 &  20 &  1.835 & 0.058 &  0.088 & 0.012  \\
              &            &                         &    &       & 667 &  1180 &  14 &  1.896 & 0.042 &  0.131 & 0.011  \\
NGC0068       & 00:18:18.7 & +30:04:17               & -3 & 14.05 & 503 &  5790 &  29 &  2.414 & 0.032 &  0.304 & 0.010  \\
NGC0078A      & 00:20:25.8 & +00:49:34               & -2 & 14.50 & 502 &  5454 &  26 &  2.398 & 0.033 &  0.308 & 0.013  \\
NGC0108       & 00:25:59.0 & +29:12:41               & -2 & 13.30 & 508 &  4786 &  25 &  2.198 & 0.038 &  0.261 & 0.012  \\
\enddata
\tablecomments{
Table 5 is presented in its entirety in the electronic edition of the 
Astronomical Journal. A portion is shown here for guidence regarding its 
form and content.
}
\end{deluxetable}

\clearpage
\begin{figure}
\centering
\mbox{\psfig{figure=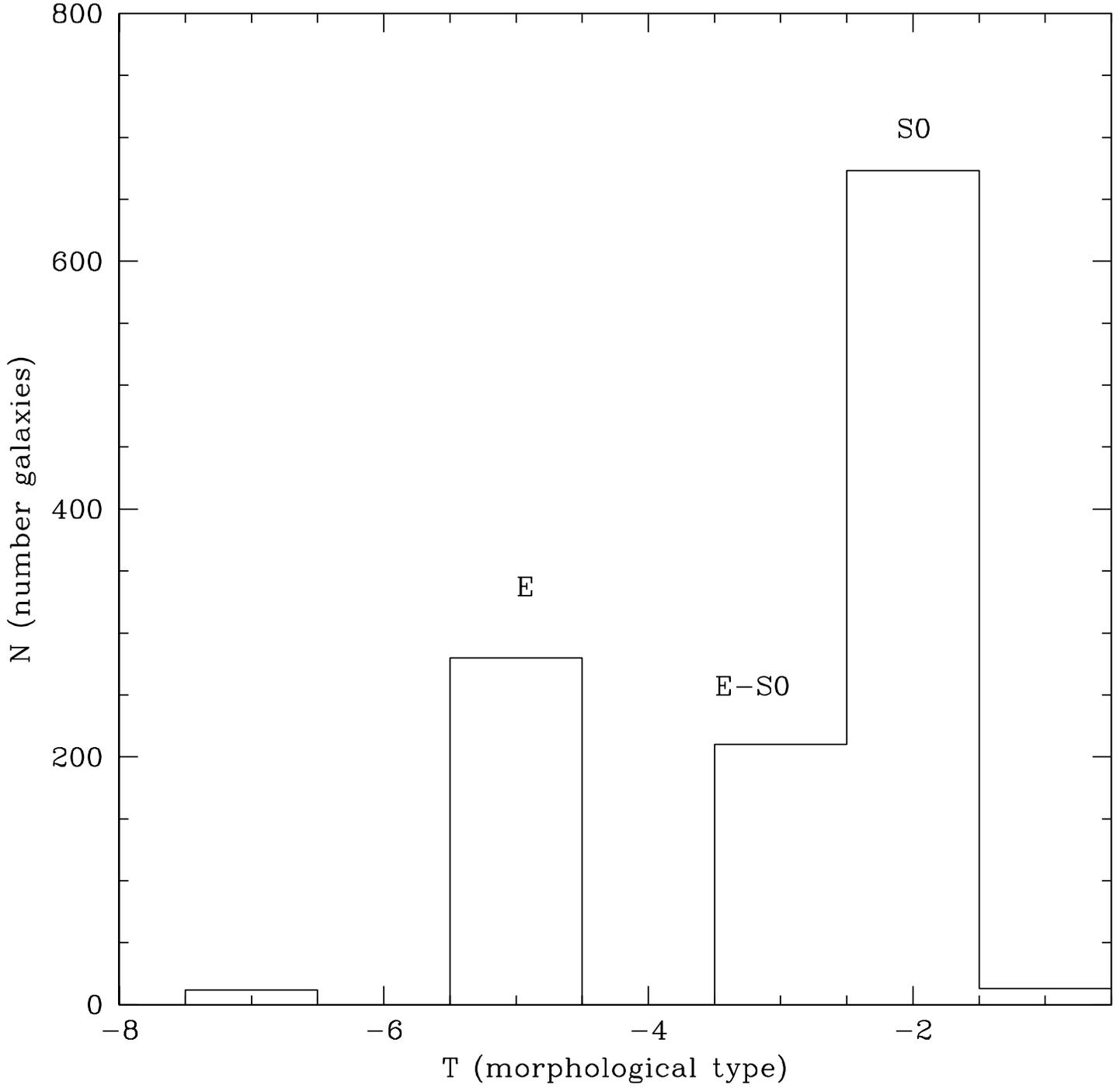,height=20truecm }}
\caption{The distribution of morphological T types on the system of
Lauberts \& Valentijn (1989) for the galaxies in this paper. 
Approximate Hubble types are also given above each of the columns.
}
\label{fig:types_ENEAR}
\end{figure}

\clearpage
\begin{figure}
\centering
\mbox{\psfig{figure=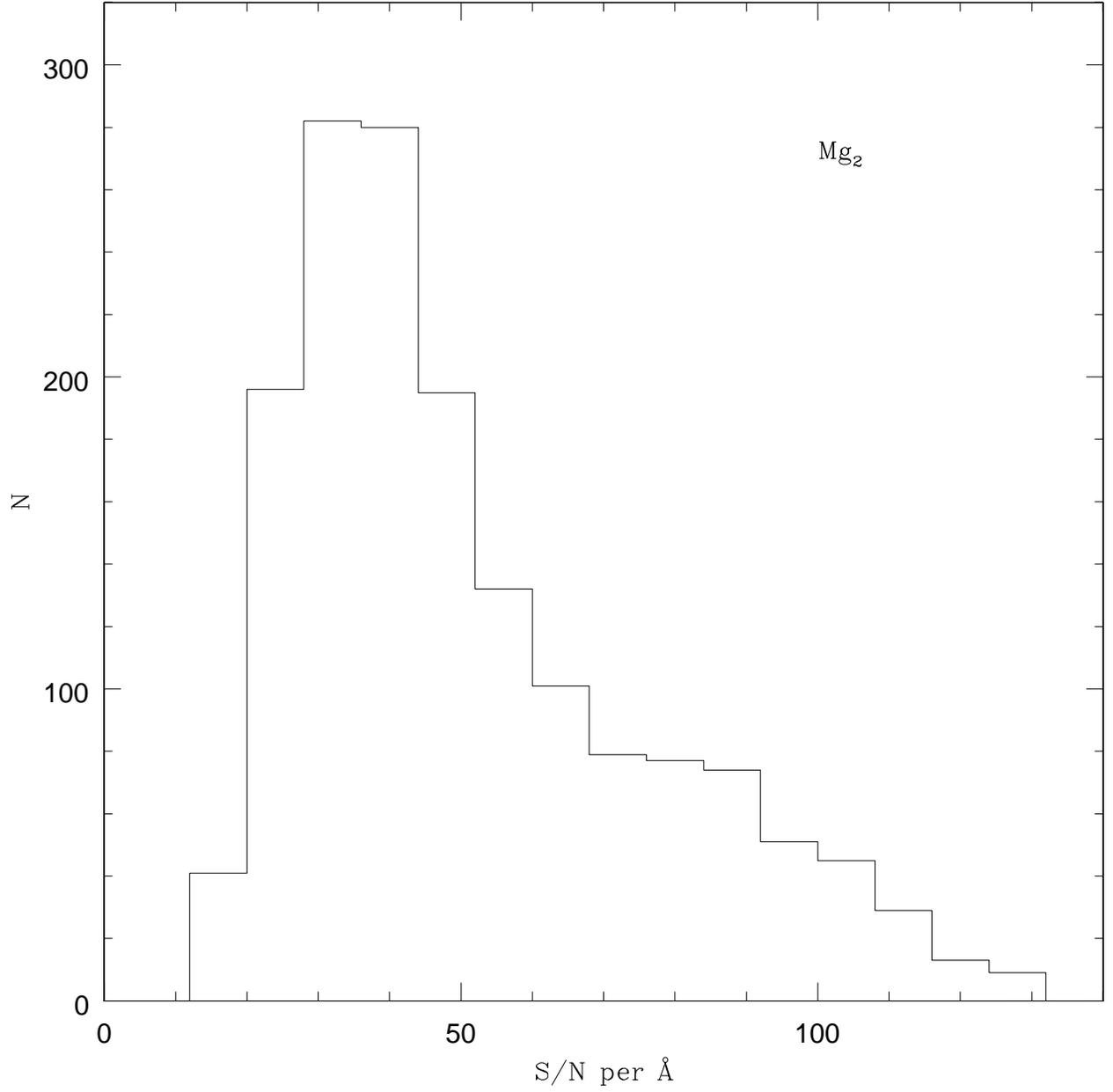,height=18truecm }}
\caption{The distribution of the $S/N$ per~\AA\, of the ENEAR spectra
in the region of the Mg$_2$ feature.}
\label{fig:SN_ENEAR}
\end{figure}

\clearpage
\begin{figure}
\centering
\mbox{\psfig{figure=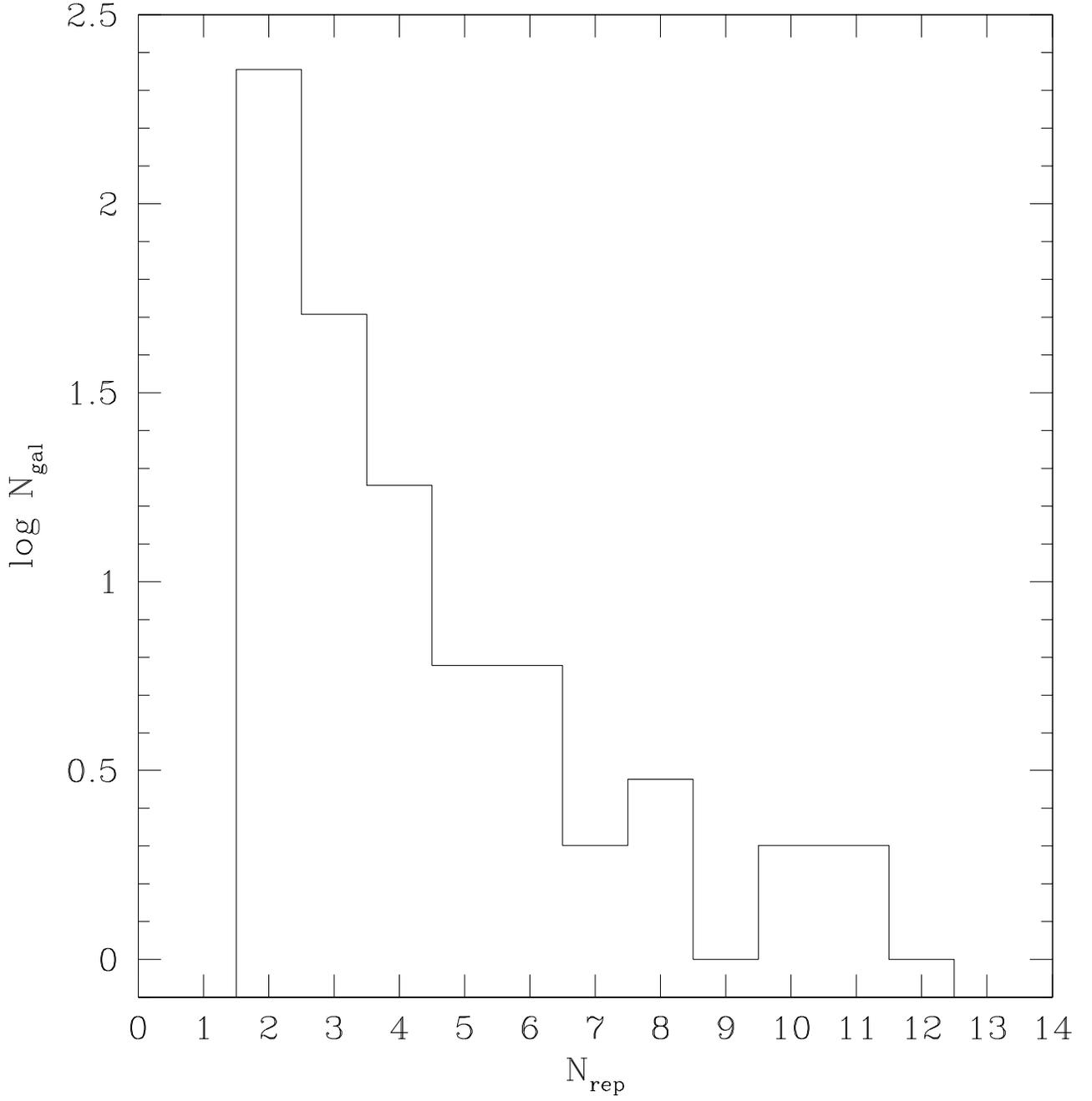,height=18truecm }}
\caption{The distribution of the internal repeated observations.}
\label{fig:histreob}
\end{figure}

\clearpage
\begin{figure}
\centering
\mbox{\psfig{figure=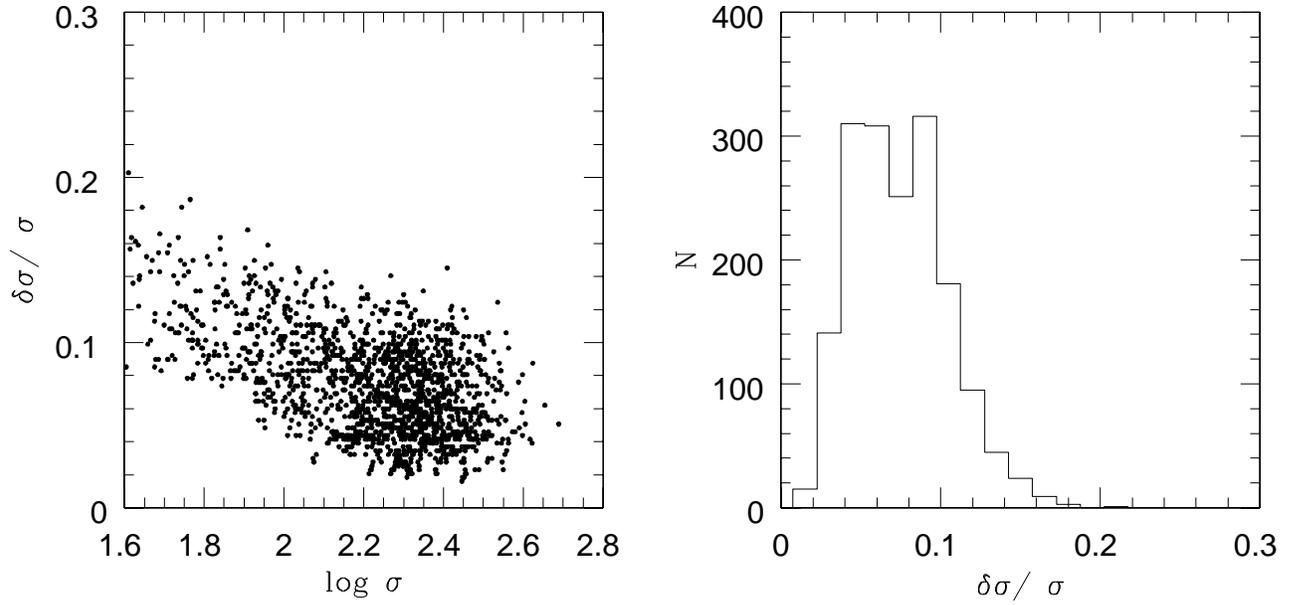,height=18truecm }}
\caption{(Left panel) The fractional error of the velocity dispersion 
$\delta~\sigma / \sigma$ as function of $\sigma$, for all the ENEAR observed spectra. (Right panel) The distribution of the velocity dispersion fractional errors $\delta~\sigma / \sigma$.}
\label{fig:sigerr}
\end{figure}

\clearpage
\begin{figure}
\centering
\mbox{\psfig{figure=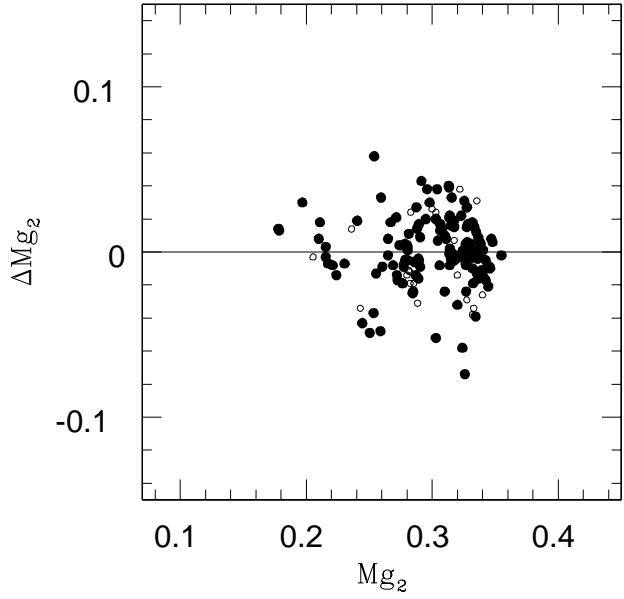,height=18truecm }}
\caption{Comparisons of the Mg$_2$ measurements
obtained by Faber \etal (1989) (Lick system) with the values
derived on the ENEAR spectra observed at low resolution (open circles),
and on ENEAR spectra observed at high resolution (filled circles).} 
\label{fig:compmg2_7S}
\end{figure}

\clearpage
\begin{figure}
\centering
\mbox{\psfig{figure=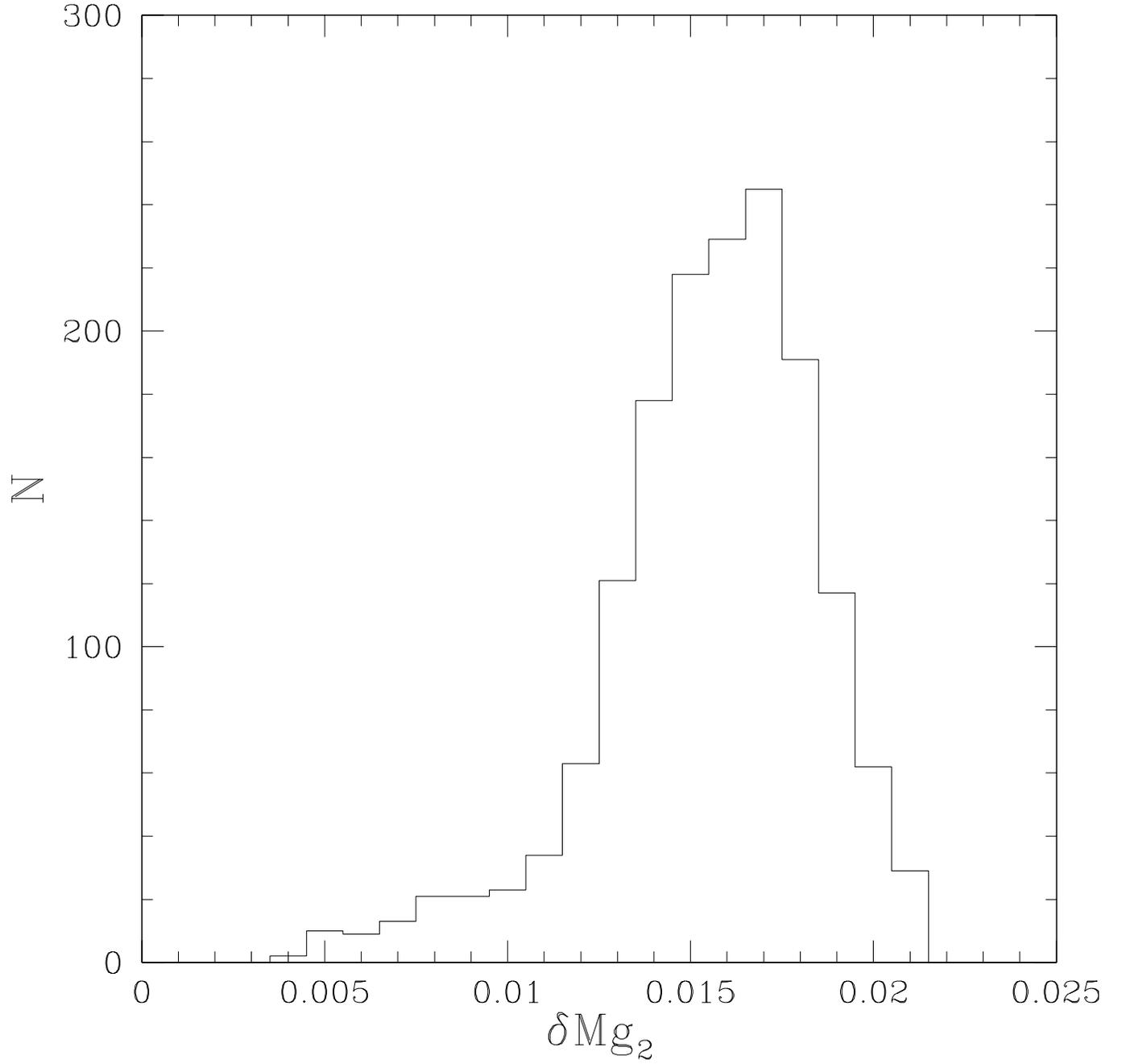,height=18truecm }}
\caption{The distribution of the errors associated to the ENEAR measurements
of the Mg$_2$ line index using the simulated spectra described in the text.}
\label{fig:histmg2err}
\end{figure}

\clearpage
\begin{figure}
\centering
\mbox{\psfig{figure=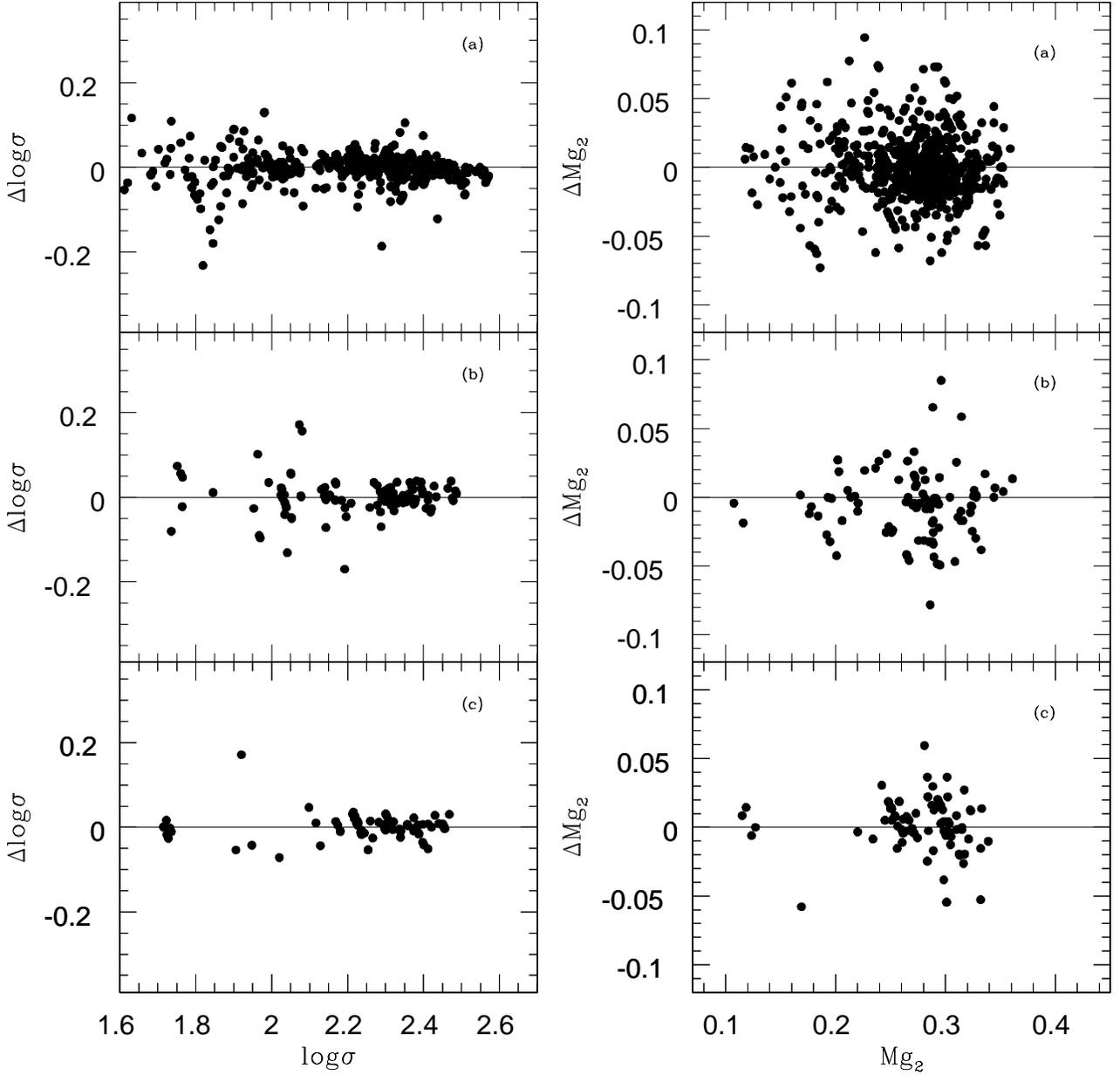,height=18truecm }}
\caption{Internal consistency of the derived velocity dispersion (left panels)
and Mg$_2$ line index (right panels). Internal comparisons between measurements
obtained at ESO (setups 1 to 4) and measurements obtained at: 
ESO (upper panels), MDM (setups 5 to 10, middle panels), and CASLEO 
(setup 11, lower panels).}
\label{fig:comparsp_int}
\end{figure}

\clearpage
\begin{figure}
\centering
\mbox{\psfig{figure=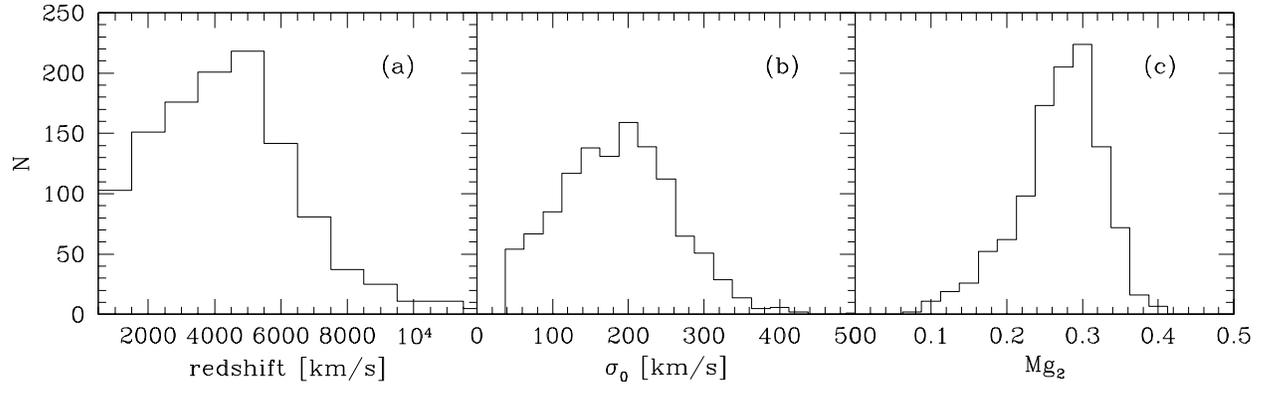,height=18truecm }}
\caption{The distribution of (a) redshift, (b) velocity dispersion, and
(c) Mg$_2$ linestrength for galaxies in the ENEAR sample.}
\label{fig:hist_sigmg2}
\end{figure}


\begin{center}
\begin{deluxetable}{lccrrcrcccccccc}
\tablecolumns{15}
\rotate
\tablewidth{0pt}
\tablenum{4}
\tabletypesize{\scriptsize}
\tablecaption{The Spectroscopic ENEAR Catalog}
\tablehead{
Name &$\alpha$ &$\delta$&                               T &  m$_{B}$& N$_{obs}$ & $cz_{hel}$ &                            $\epsilon_{cz_{hel}}$  & $\log_{10} \sigma$ &                  $\epsilon_{\log_{10} \sigma}$& N$_{Mg_2}$ & Mg$_2$ &           $\epsilon_{\rm{Mg}_2}$& Notes & Lit\\
 & (2000) & (2000) & & mag &  & $km s^{-1}$ &             $km s^{-1}$ &$km s^{-1}$ &$km s^{-1}$ & &  & & & \\
\tablevspace{5pt}
(1) & (2) & (3) & (4) & (5) & (6) & (7) & (8)             & (9) & (10) & (11) &                                             (12) & (13) & (14)  & (15)}
\startdata
ESO409G012    & 00:04:42.2 & -30:29:00& -5 & 14.23 &  1 &  8043 &  63 &  2.386 & 0.040 &  1 &  0.242 & 0.015 & 1 &   *  \\      
IC1529        & 00:05:13.3 & -11:30:12& -2 & 14.50 &  3 &  6726 &  25 &  2.258 & 0.022 &  3 &  0.255 & 0.014 & 1 &   \, \\      
NGC7832       & 00:06:28.4 & -03:42:58& -3 & 13.50 &  2 &  6202 &  19 &  2.351 & 0.032 &  2 &  0.285 & 0.016 & 1 &   \, \\      
UGC00061      & 00:07:23.8 & +47:02:26& -2 & 14.30 &  2 &  5354 &  52 &  2.312 & 0.031 &  2 &  0.303 & 0.013 & 1 &   \, \\      
NGC0043       & 00:13:00.8 & +30:54:55& -2 & 13.90 &  2 &  4846 &  20 &  2.298 & 0.042 &  2 &  0.323 & 0.013 & 1 &   \, \\      
UGC00130      & 00:13:56.9 & +30:52:58& -7 & 14.20 &  1 &  4791 &  30 &  2.146 & 0.042 &  1 &  0.273 & 0.017 & 1 &   \, \\      
NGC0050       & 00:14:44.5 & -07:20:38& -3 & 12.50 &  1 &  5468 &  22 &  2.422 & 0.026 &  0 &  0.000 & 0.000 & 1 &   \, \\      
NGC0063       & 00:17:45.6 & +11:27:01& -5 & 12.60 &  2 &  1167 &  26 &  1.875 & 0.060 &  1 &  0.131 & 0.020 & 4, 3 &   \, \\
\enddata
\tablecomments{(1)  no problems;
(2) star along the slit;
(3) emission lines;
(4) low S/N measurements;
(5) low velocity dispersion on the limit of the resolution;
(6) old data (Reticon);
(7) peculiar spectrum: eg. broad lines (supernova?),
absorption lines too weak or undetectable.
}
\end{deluxetable}
\end{center}
\begin{center}
\begin{deluxetable}{lccrrcrccccc}
\tablecolumns{12}
\rotate
\tablewidth{0pt}
\tablenum{5}
\tabletypesize{\scriptsize}
\tablecaption{Individual ENEAR Spectroscopic          Measurements}
\tablehead{
Name &$\alpha$ &$\delta$&                               T &  m$_{B}$& Run & $cz_{hel}$ &                                  $\epsilon_{cz_{hel}}$  & $\log_{10} \sigma$ &                  $\epsilon_{\log_{10}\sigma}$& Mg$_2$ &                         $\epsilon_{\rm{Mg}_2}$ \\ 
 & (2000) & (2000) & & mag &  & $km s^{-1}$ &             $km s^{-1}$ &$km s^{-1}$ &$km s^{-1}$ & mag & mag \\ 
\tablevspace{5pt}
(1) & (2) & (3) & (4) & (5) & (6) & (7) & (8)             & (9) & (10) & (11) &                                             (12) }
\startdata
ESO409G012    & 00:04:42.2 & -30:29:00               & -5 & 14.23 & 660 &  8044 &  63 &  2.386 & 0.040 &  0.242 & 0.015  \\
IC1529        & 00:05:13.3 & -11:30:12               & -2 & 14.50 & 653 &  6735 &  30 &  2.251 & 0.029 &  0.257 & 0.014  \\
              &            &                         &    &       & 656 &  6725 &  22 &  2.259 & 0.018 &  0.248 & 0.009  \\
              &            &                         &    &       & 653 &  6724 &  23 &  2.260 & 0.016 &  0.270 & 0.014  \\
NGC7832       & 00:06:28.4 & -03:42:58               & -3 & 13.50 & 653 &  6202 &  22 &  2.368 & 0.034 &  0.276 & 0.016  \\
              &            &                         &    &       & 653 &  6203 &  16 &  2.343 & 0.024 &  0.290 & 0.012  \\
UGC00061      & 00:07:23.8 & +47:02:26               & -2 & 14.30 & 508 &  5362 &  57 &  2.311 & 0.036 &  0.296 & 0.016  \\
              &            &                         &    &       & 508 &  5350 &  47 &  2.312 & 0.024 &  0.304 & 0.007  \\
\enddata
\end{deluxetable}
\end{center}
\end{document}